 \documentclass[acmlarge,screen, nonacm]{acmart}

\AtBeginDocument{%
  \providecommand\BibTeX{{%
    \normalfont B\kern-0.5em{\scshape i\kern-0.25em b}\kern-0.8em\TeX}}}

\setcopyright{acmlicensed}
\copyrightyear{2024}
\acmYear{2024}
\acmDOI{XXXXXXX.XXXXXXX}

\acmConference[CHI EA '24]{Workshop on Shaping The Future: Developing Principles for Policy Recommendations for Responsible Innovation in Virtual Worlds}{May 11--16,
2024}{Honolulu, HI, USA}

\begin{document}

\title{Exploring Vulnerabilities in Remote VR User Studies}

\author{Viktorija Paneva}
\orcid{0000-0002-5152-3077}
\author{Florian Alt}
\orcid{0000-0001-8354-2195}
\affiliation{%
  \institution{University of the Bundeswehr Munich}
  \city{Munich}
  \country{Germany}
}




\begin{abstract}
This position paper explores the possibilities and challenges of using Virtual Reality (VR) in remote user studies. 
Highlighting the immersive nature of VR, the paper identifies key vulnerabilities, including varying technical proficiency, privacy concerns, ethical considerations, and data security risks. 
To address these issues, proposed mitigation strategies encompass comprehensive onboarding, prioritized informed consent, implementing privacy-by-design principles, and adherence to ethical guidelines. 
Secure data handling, including encryption and disposal protocols, is advocated. 
In conclusion, while remote VR studies present unique opportunities, carefully considering and implementing mitigation strategies is essential to uphold reliability, ethical integrity, and security, ensuring responsible and effective use of VR in user research. Ongoing efforts are crucial for adapting to the evolving landscape of VR technology in user studies.
\end{abstract}





\maketitle

\section{Introduction}

In recent years, Virtual Reality (VR) has been identified as a powerful tool for user studies due to its ability of delivering immersive experiences that can replicate real-world scenarios. 
The flexibility of VR extends to simulating and effortlessly switching between various interfaces and interactions~\cite{Jetter2020}. 
Furthermore, it allows for the development of digital twins of physical systems, representing a significant stride in user study methodologies in HCI, especially in the context of evaluating physical user interfaces.
Recent user studies have begun exploring the opportunities of evaluating user interfaces using VR simulations. 
Examples include assessing the feasibility of evaluating physical user interfaces in VR by incorporating a built-in simulation module of the physical dynamics~\cite{Paneva2020}, virtual field studies on public displays~\cite{Maekelae2020} and an exploration of shoulder surfing behavior \cite{abdrabou2022avi}. 

Another advantage of remote VR studies is the possibility of engaging a more diverse and representative participant pool across different geographical regions. 
This inclusivity encompasses individuals working from home, those involved in childcare, or those facing mobility challenges -- demographics frequently overlooked in conventional studies. 
Conducting studies in participants' homes not only enhances comfort but also simplifies logistical considerations.
A comprehensive exploration of opportunities and challenges related to remote user studies in VR is detailed in~\cite{Radiah2021}.

However, the shift towards immersive technologies packed with sensors introduces a set of vulnerabilities that necessitate careful consideration~\cite{Spiegel2018, Abraham2022, Nair2023}. 
This position paper explores some of the most prevalent challenges inherent to remote user studies in VR and proposes strategic approaches to mitigate these vulnerabilities.

 
\section{Key Vulnerabilities}
We consider the following vulnerabilities to require attention when running remote VR studies.

\textbf{Technical Challenges}
While the shift towards remote participation removes geographical barriers, it introduces a new concern -- a participant pool with varying levels of technical literacy and experience. 
This diversity may encompass individuals with limited familiarity with technology, requiring thoughtful consideration and planning to ensure a fair and inclusive study that adheres to all data privacy and security standards.


\textbf{Privacy Concerns}
The immersive nature and extensive tracking involved in VR raise the risk of unintentional exposure of personal information.
Data collected from motion sensors, eye-tracking devices, and physiological monitoring could reveal details regarding health conditions, disabilities, and the user's emotional or mental well-being. 
Furthermore, tracking extends to the user's surroundings, including people and objects in the vicinity.

\textbf{Ethical Considerations}
In remote studies, the absence of direct supervision could pose a challenge to upholding the ethical guidelines (i.e., ensuring the physical safety of participants).
This gap may allow participants to potentially engage in harmful activities, with researchers facing difficulties in promptly intervening. 

\textbf{Data Security}
The transmission and remote storage of user data introduce significant concerns regarding data security, such as interception risk and unauthorized access to potentially sensitive information.
This includes participant interactions and biometric data, which, if compromised, could be exploited for activities like avatar impersonation and virtual identity theft.

\section{Mitigation Strategies}
The aforementioned key vulnerabilities could be addressed through the following strategies.

\textbf{Technical Inclusivity}
Create user-friendly interfaces and a comprehensive onboarding process. Develop intuitive VR platforms with clear instructions to minimize the technical learning curve. 
Consider strategies like remote technical assistance during studies and integrating training modules to help participants acquire the necessary technical competencies quickly.

\textbf{Privacy Safeguards}
Prioritise informed consent that clearly communicates the scope of data collection and intended use. 
Devise comprehensive and easily understandable consent forms outlining potential risks and mitigation measures.
Use anonymization techniques to safeguard user identities, exercise responsible use of recording features, and, where applicable, aggregate data to reduce the risk of individual identification.
Adopt privacy-by-design principles, incorporating privacy considerations from the initial stages of the experimental design and continuously assessing and improving privacy features throughout the iterative development process.

\textbf{Ethical Guidelines}
Develop and enforce strict ethical guidelines tailored for remote user studies in VR. Researchers should conduct thorough briefings with participants, emphasizing responsible use of the technology and establishing mechanisms for immediate intervention in case of ethical concerns.

\textbf{Secure Data Handling}
Employ secure communication channels for data transmission and storage, using encryption to safeguard sensitive information.  
Establish well-defined protocols for post-study data disposal, outlining secure deletion or anonymization procedures to mitigate the risk of unintended exposure or misuse.
Regulatory compliance, adherence to data protection laws, and developing industry standards for VR privacy (e.g., a code of ethics for developers~\cite{Adams2018}) are also important steps to ensure a consistent and secure user environment. 

\section{Conclusion}
While remote user studies in VR offer unique opportunities, they have inherent vulnerabilities that demand careful attention. By addressing technical challenges, privacy concerns, ethical considerations, and data security issues, researchers can enhance remote VR studies' reliability and ethical integrity. As the field continues to evolve, ongoing efforts to identify and mitigate vulnerabilities will be crucial in ensuring the responsible and effective use of VR technologies for user research.


\bibliographystyle{ACM-Reference-Format}
\bibliography{sample-base}


\end{document}